\documentclass{svmult}

\usepackage{graphicx}    

\def\beq{\begin{equation}}
\def\eeq{\end{equation}}

\def\bea{\begin{eqnarray}}
\def\eea{\end{eqnarray}}
\def\ba{\begin{array}}                  
\def\ea{\end{array}}

\newcommand{\pam}{\partial_\mu}
\newcommand{\pan}{\partial_\nu}
\newcommand{\parh}{\partial_\rho}

\begin{document}

\title*{The Noncommutative Standard Model and Forbidden Decays}
\author{Peter Schupp\inst{1}\and
Josip Trampeti\'c\inst{2}}
\institute{International University Bremen, Campus Ring 8, 28759 Bremen, Germany
\texttt{p.schupp@iu-bremen.de}
\and Theoretical Physics Division, Rudjer Bo\v skovi\'c Institute, 10002 Zagreb, Croatia \texttt{josipt@rex.irb.hr}}

\maketitle

\section{Introduction}

In this contribution we discuss the
Noncommutative Standard Model and the associated Standard
Model-forbidden decays that can possibly serve as an experimental signature
of space-time noncommutativity.

The idea of quantized space-time and noncommutative field theory has a
long history that can be traced back to Heisenberg \cite{Heisenberg} and Snyder \cite{sny}.
A noncommutative structure of spacetime can be introduced by
promoting the usual spacetime 
coordinates $x$ to noncommutative (NC) coordinates 
$\hat x$ with
\begin{equation}
\left[{\hat x}^{\mu},{\hat x}^{\nu} \right]=i\theta^{\mu\nu},                \label{CR}
\end{equation}
were $\theta^{\mu\nu}$ is a real antisymmetric matrix.
A noncommutativity scale $\Lambda_{NC}$ is fixed
by choosing dimensionless matrix elements 
$c^{\mu\nu}=\Lambda_{NC}^2\,
\theta^{\mu\nu}$ of order one.
The original motivation to study such a scenario
was the hope that the introduction
of a fundamental scale
could deal with the infinities of
quantum field theory in a natural way \cite{Filk}.
The mathematical theory that replaces
ordinary differential geometry in the description of quantized
spacetime is noncommutative geometry \cite{connes2}. 
A realization of
the electroweak sector of the Standard Model in the framework of noncommutative
geometry can be found \cite{connes1}, where the Higgs field
plays the role of a gauge boson in the non-commutative (discrete)
direction. This model is noncommutative in an extra internal direction
but not in spacetime itself. It is therefore not the focus of the
present work, although it can in principle be combined with it.

Noncommutativity of spacetime
is very natural in string theory and can be
understood as an effect of the interplay of closed and open strings.
The commutation relation (\ref{CR}) 
enters in string theory through
the Moyal-Weyl star product
\begin{equation}
f \star g = \sum_{n=0}^\infty \frac{\theta^{\mu_1 \nu_1} 
\cdots \theta^{\mu_n \nu_n}}{(-2i)^nn!}  \;
\partial_{\mu_1}\ldots\partial_{\mu_n} f
\cdot\partial_{\nu_1}\ldots\partial_{\nu_n} g.
\end{equation} 
For coordinate functions: $x^\mu \star x^\nu - x^\nu \star x^\mu = i \theta^{\mu\nu}$.
The tensor $\theta^{\mu\nu}$ is determined by a NS  $B^{\mu\nu}$-field and
the open string metric $G^{\mu\nu}$~\cite{DH}, which both
depend on a given closed string background. 
The effective physics on D-branes is most naturally captured by
noncommutative gauge theory, but it can also be described by
ordinary gauge theory. Both descriptions
are related by the Seiberg-Witten (SW) map~\cite{SW}, which 
expresses noncommutative gauge fields in terms of fields
with ordinary ``commutative'' gauge transformation properties.

The star product formalism in conjunction with the Seiberg-Witten map
of fields naturally leads to a perturbative
approach to field theory on noncommutative spaces.
It is particulary well-suited to study Standard Model-forbidden
processes induced by spacetime noncommutativity. This formalism 
can also be used to study non-perturbative noncommutative effects.
 In particular cases an algebraic approach may be more
convenient for actual computations but the structure of the star product 
results can still be a useful guideline.

A method for implementing non-Abelian $SU(N)$ Yang-Mills theories on
non-commu\-ta\-tive spacetime has been proposed in
\cite{Madore:2000en,Jurco:2000ja,Jurco:2001my,Jurco:2001rq}. 
  In \cite{Calmet:2001na} this method has
been applied to the
full Standard Model of particle physics \cite{Glashow:1961tr} resulting
in a minimal non-commutative extension of the Standard Model with structure 
group $SU(3)_C \times SU(2)_L \times U(1)_Y$ and with the same
fields and the same number of coupling
parameters as in the original Standard Model.
It is the only known approach that allows to build models of the
electroweak sector directly based on the structure group $SU(2)_L \times U(1)_Y$
in a noncommutative background.
Previously only $U(N)$ gauge theories were under control, and it was thus only possible to consider
extensions of the Standard Model. Furthermore there were problems with
the allowed charges and with the gauge invariance
of the Yukawa terms in the action.

In an alternative approach to the construction of a noncommutative
generalization of the Standard Model
the usual
problems of noncommutative model buildings, i.e., charge quantization and the restriction of
the noncommutative gauge group are circumvented by
enlarging the gauge group to $U(3)\times U(2) \times U(1)$ \cite{Chaichian:2001py}. 
The hypercharges and the
electric charges are  quantized to the correct values of the
usual quarks and leptons, however, there are some open issues with the NC
gauge invariance of the Yukawa terms. In principle the two approaches can be
combined.

\section{The Noncommutative Standard Model}

\subsection{Noncommutative Yang-Mills}

Consider an ordinary  Yang-Mills action with  gauge group $G$, 
where $G$ is a compact simple Lie group, and a fermion multiplet $\Psi$
\beq
S=\int d^4x \,{-1\,\over 2 g^2}^{\,} 
\mathrm{Tr}(F_{\mu\nu}F^{\mu\nu})  + 
\overline{\Psi}  i
{/\!\!\!\!D} \Psi \label{A1m}
\eeq
This action is gauge invariant under 
\beq
\delta\Psi=i\rho_\Psi(\Lambda)\Psi\label{psi}
\eeq
where $\rho_{\Psi}$ is the representation of $G$  determined by the
multiplet $\Psi$.
The noncommutative generalization of 
(\ref{A1m}) is given by 
\beq
\widehat S =\int d^4x \,{-1\,\over 2 g^2}^{\,} 
T^{\!}r(\widehat F_{\mu\nu}\*\widehat F^{\mu\nu})  + 
\overline{\widehat \Psi} \star i
\widehat{/\!\!\!\!D} \widehat \Psi \label{Action1multiplet}
\eeq
where the noncommutative field strength $\widehat F$ is defined by
\beq
\widehat F_{\mu\nu}  = \pam\widehat A_\nu 
- \pan\widehat A_\mu  -i[\widehat A_\mu,\widehat A_\nu]_\star.
\eeq
The covariant derivative is given by
\beq
\widehat D_\mu \widehat\Psi = \pam \widehat\Psi 
- i \rho_\Psi(\widehat A_\mu)\star\widehat\Psi\,.
\label{covder}
\eeq
The action (\ref{Action1multiplet}) is invariant under the 
noncommutative gauge transformations 
\beq
\hat\delta\widehat\Psi = i \rho_\Psi(\widehat\Lambda) \star
\widehat\Psi , \qquad
\hat\delta \widehat A_\mu = \pam\widehat\Lambda 
+ i[\widehat\Lambda,\widehat A_\mu]_\star,
\qquad \hat\delta\widehat F_{\mu\nu} 
= i[\widehat\Lambda,\widehat F_{\mu\nu}]_\star~.
\label{deltaA}
\eeq
If the gauge fields are assumed to be Lie-algebra valued, it appears
that only $U(N)$ in the fundamental representation is consistent with
noncommutative gauge transformations: Only in this case the commutator
\beq
[\widehat\Lambda,\widehat\Lambda']_\star
= \frac{1}{2}\{\Lambda_a(x)\stackrel{\star}{,}
\Lambda'_b(x)\}[T^a,T^b] + \frac{1}{2}[\Lambda_a(x)\stackrel{\star}{,}\Lambda'_b(x)]\{T^a,T^b\}
\label{com}
\eeq
of two Lie algebra-valued non-commutative gauge parameters 
$\widehat\Lambda = \Lambda_a(x) T^a$ and $\widehat\Lambda' = \Lambda'_a(x) T^a$
again closes in the Lie algebra \cite{Madore:2000en,Jurco:2000ja}. 
The fact that a $U(1)$ factor cannot easily be decoupled from NC $U(N)$, can also be
seen by noting the interactions of $SU(N)$ gluons and $U(1)$ (hyper) photons in NC 
Yang-Mills theory \cite{Armoni:2000xr}. 
For a sensible phenomenology of particle physics on
noncommutative spacetime we need to be able to use other gauge groups.
Furthermore, in the special case of $U(1)$ a similar argument
show that charges are quantized to values $\pm e$ and zero.
These restrictions can be avoided if we allow gauge fields and gauge transformation
parameters that are
valued in the enveloping algebra of the gauge group. 
\beq
\widehat\Lambda = \Lambda^0_a(x) T^a +  \Lambda^1_{ab}(x) T^a T^b
+ \Lambda^2_{abc}(x) T^a T^b T^c + \ldots
\eeq
A priori we now face the problem
that we have an infinite number of parameters
$\Lambda^0_a(x)$, $\Lambda^1_{ab}(x)$, $\Lambda^2_{abc}(x)$, \ldots,
but these are not independent. They can in fact
all be expressed in terms of the right number of classical
parameters and fields via the Seiberg-Witten maps.
The non-commutative fields $\widehat A$, $\widehat \Psi$ and 
non-commutative gauge parameter $\widehat\Lambda$ can be expressed
as ``towers'' built upon the corresponding ordinary fields $A$, $\Psi$  
and ordinary gauge parameter $\Lambda$. The Seiberg-Witten
maps \cite{Seiberg:1999vs} express non-commutative fields and parameters as local 
functions of the ordinary fields and parameters,
\begin{eqnarray}
\widehat A_\xi[A] & = & A_\xi 
+ \frac{1}{4} \theta^{\mu\nu}\{A_\nu,\pam A_\xi\} + 
\frac{1}{4} \theta^{\mu\nu}\{F_{\mu\xi},A_\nu\} +  \mathcal{O}(\theta^2)
\label{SWA}\\
\widehat \Psi[\Psi,A] & = & \Psi 
+ \frac{1}{2} \theta^{\mu\nu}\rho_\Psi(A_\nu)\pam\Psi
+\frac{i}{8}\theta^{\mu\nu}[\rho_\Psi(A_\mu), \rho_\Psi(A_\nu)] \Psi + \mathcal{O}(\theta^2)
 \label{SWPsi}\\
\widehat\Lambda[\Lambda, A] & = & \Lambda 
+ \frac{1}{4} \theta^{\mu\nu}\{A_\nu,\pam \Lambda\}+ \mathcal{O}(\theta^2)
\label{SWLambda}
\end{eqnarray}
where $F_{\mu\nu} = \pam A_\nu - \pan A_\mu -i [A_\mu,A_\nu]$ is the
ordinary field strength.
The Seiberg-Witten maps have the remarkable property that ordinary gauge
transformations $\delta A_\mu = \pam \Lambda + i[\Lambda,A_\mu]$
and $\delta \Psi = i \Lambda\cdot \Psi$ 
induce non-commutative gauge transformations
(\ref{deltaA}) of the fields 
$\widehat A$, $\widehat \Psi$ with gauge parameter $\widehat \Lambda$.

\subsection{Standard model fields}

The Standard Model gauge group
is $G_{SM}=SU(3)_C \times SU(2)_L \times U(1)_Y$.
The gauge potential $A_\mu$ and gauge 
parameter ${\Lambda}$ are valued in Lie($G_{SM}$):
\begin{eqnarray}
 {A_\nu}&=&g' {\cal A}_\nu(x)Y+g \sum_{a=1}^{3} B_{\nu a}(x) T^a_L
  +g_S \sum_{b=1}^{8} G_{\nu b}(x) T^b_S 
  \\
  {\Lambda}&=&g' \alpha (x)Y+g \sum_{a=1}^{3} \alpha^L_{a}(x) T^a_L
  +g_S \sum_{b=1}^{8} \alpha^S_{b}(x) T^b_S,
\end{eqnarray}
where $Y$, $T^a_L$, $T^b_S$ are the generators of
$u(1)_Y$, $su(2)_L$ and $su(3)_C$ respectively.
In addition to the gauge bosons we have
three families of left- and right-handed fermions and a Higgs doublet
\beq
\Psi^{(i)}_L= \left ( \matrix{ L^{(i)}_L \cr
 Q^{(i)}_L
    } \right),\qquad
 \Psi^{(i)}_R = \left ( \matrix{ e^{(i)}_R \cr
 u^{(i)}_R \cr   d^{(i)}_R
    } \right), \qquad 
    {\Phi} =
 \left(\begin{array}{c}  \phi^+ \\  \phi^0
   \end{array} \right )
\eeq 
where i = 1,2,3 is the generation index and
$\phi^+$,
$\phi^0$ are complex scalar fields. We shall
now apply the appropriate 
SW maps to the fields $A_\mu$, $\Psi^{(i)}$, $\Phi$, expand to first
order in $\theta$ and write the corresponding
NC Yang-Mills action \cite{Calmet:2001na}.

\subsection{Noncommutative Yukawa terms}

Special care must be taken in the definition
of the trace in the gauge kinetic terms and in the construction
of covariant Yukawa terms.
The classical Higgs field $\Phi(x)$ commutes with the generators of
the $U(1)$ and $SU(3)$ gauge transformations. It also commutes
with the corresponding gauge parameters. The latter is no longer
true  in the noncommutative setting:
The coefficients $\alpha(x)$ and $\alpha_b^S(x)$ of
the $U(1)$ and $SU(3)$ generators in the gauge parameter
are functions and therefore do not $\star$-commute 
with the Higgs field.
This makes it hard to write down covariant Yukawa terms. The solution to
the problem is the hybrid SW map \cite{Schupp:2001we}
\bea
\widehat\Phi[\Phi,A,A'] = &&\Phi + \frac{1}{2}\theta^{\mu\nu} A_\nu
\Big(\pam\Phi -\frac{i}{2} (A_\mu \Phi - \Phi A'_\mu)\Big) \nonumber\\
&&+\frac{1}{2}\theta^{\mu\nu} 
\Big(\pam\Phi -\frac{i}{2} (A_\mu \Phi - \Phi A'_\mu)\Big)A'_\nu
+ \mathcal{O}(\theta^2)\label{SWPhi} .
\eea
By choosing appropriate
representations it allows us to assign
separate left and right charges to the noncommutative Higgs field
$\widehat\Phi$ that add up to its usual charge \cite{Calmet:2001na}. 
Here are two examples:
\beq
\begin{array}{rccccccccccc}
 & \overline{\widehat L}_L &\star &\rho_L(\widehat\Phi) &\star &\widehat e_R & \:\qquad&
\overline{\widehat Q}_L &\star &\rho_Q(\widehat\Phi) &\star &\widehat d_R
\\[1ex]
{ Y = \quad} & 1/2& &\underbrace{{ -1/2 } + { 1}}_{1/2} && -1 &\:\qquad&
-1/6& &\underbrace{{ 1/6} + { 1/3}}_{1/2} && -1/3 
\end{array}
\eeq 
We see here two instances of a general rule: The gauge fields in the SW maps
and in the covariant derivatives inherit
their representation (charge for $Y$, trivial or fundamental 
representation for $T^a_L$, $T^b_S$)
from the fermion fields $\Psi^{(i)}$ to their left and to their right.

In GUTs it is more natural to first combine the left-handed and right-handed
fermion fields and then contract the resulting expression
with Higgs fields to obtain a gauge invariant Yukawa term.
Consequently in NC GUTs we need to use the hybrid SW map for the left-handed
fermion fields and then sandwich them between the NC Higgs on the left and
the right-handed fermion fields on the right  \cite{Aschieri:2002mc}.

\subsection{The minimal NCSM}

The trace  in the kinetic terms for the gauge bosons is not unique,
it depends on the choice of representation.
This would not matter if the gauge fields
were Lie algebra valued, but in the noncommutative case
they live in the enveloping algebra.
The simplest choice is a sum of three
traces over the $U(1)$, $SU(2)$, $SU(3)$ sectors
with $Y = \frac{1}{2} \Big({1 \, \phantom{-}0 \atop 0  \,-1}\Big)$
in the definition of $\mbox{{\bf tr}}_{\bf 1}$ and the
fundamental representation for $\mbox{{\bf tr}}_{\bf 2}$
and $\mbox{{\bf tr}}_{\bf 3}$.
This leads to the following gauge kinetic terms
\begin{eqnarray}
  S_{\mathrm{gauge}}&=&-\frac{1}{4} \, \int d^4x \, f_{\mu \nu} f^{ \mu \nu}
-\frac{1}{2} \, {\rm Tr} \int d^4x \, F^L_{\mu \nu} F^{L \mu \nu}
\nonumber \\ \nonumber &&
-\frac{1}{2} \, {\rm Tr}  \int d^4x \, F^S_{\mu \nu} F^{S \mu \nu}
+ \frac{1}{4}g_S \, \theta^{\mu \nu} \, {\rm Tr} \int d^4x \, F^S_{\mu \nu}
F^S_{\rho \sigma} F^{S \rho \sigma}\\
&&  - g_S \, \theta^{\mu
\nu} \, {\rm Tr} \int d^4x \, F^S_{\mu \rho} F^S_{\nu \sigma} F^{S \rho
\sigma} + {\cal O}(\theta^2)\;.
\end{eqnarray}
Note, that there are no new triple $f$ or 
triple $F^L$-terms.

The full action of the Minimal Noncommutative
Standard Model is \cite{Calmet:2001na}:
\begin{eqnarray}
S_{NCSM}&=&\int d^4x \sum_{i=1}^3 \overline{\widehat \Psi}^{(i)}_L \star i
\widehat{/\!\!\!\! D} \widehat \Psi^{(i)}_L
+\int d^4x \sum_{i=1}^3 \overline{\widehat \Psi}^{(i)}_R \star i
\widehat{/\!\!\!\!  D} \widehat \Psi^{(i)}_R \nonumber \\ 
&-& \int d^4x \frac{1}{2 g'} 
\mbox{{\bf tr}}_{\bf 1} \widehat
F_{\mu \nu} \star  \widehat F^{\mu \nu}
-\int d^4x \frac{1}{2 g} \mbox{{\bf tr}}_{\bf 2} \widehat
F_{\mu \nu} \star  \widehat F^{\mu \nu}\nonumber \\
&-& \int d^4x \frac{1}{2 g_S} \mbox{{\bf tr}}_{\bf 3} \widehat
F_{\mu \nu} \star  \widehat F^{\mu \nu}
+ \int d^4x \Bigg( \rho_0(\widehat D_\mu \widehat \Phi)^\dagger
\star \rho_0(\widehat D^\mu \widehat \Phi)            
\nonumber \\ && 
- \mu^2 \rho_0(\widehat {\Phi})^\dagger \star  \rho_0(\widehat \Phi) - \lambda
\rho_0(\widehat \Phi)^\dagger \star  \rho_0(\widehat \Phi)
\star
\rho_0(\widehat \Phi)^\dagger \star  \rho_0(\widehat \Phi)   \Bigg)
\nonumber \\ 
&-& \int d^4x \left (\sum_{i,j=1}^3 W^{ij} \Big
( ( \bar{ \widehat L}^{(i)}_L \star \rho_L(\widehat \Phi))
\star  \widehat e^{(j)}_R
+ \bar {\widehat e}^{(i)}_R \star (\rho_L(\widehat \Phi)^\dagger \star \widehat
L^{(j)}_L) \Big ) \right.
\nonumber \\ && 
+\sum_{i,j=1}^3 G_u^{ij} \Big
( ( \bar{\widehat Q}^{(i)}_L \star \rho_{\bar Q}(\widehat{\bar\Phi}))\star  
\widehat u^{(j)}_R
+ \bar {\widehat u}^{(i)}_R \star 
(\rho_{\bar Q}(\widehat{\bar\Phi})^\dagger
\star \widehat Q^{(j)}_L) \Big )
\nonumber \\ &&  \left.
+\sum_{i,j=1}^3 G_d^{ij} \Big
( ( \bar{ \widehat Q}^{(i)}_L \star \rho_Q(\widehat \Phi))\star  
\widehat d^{(j)}_R
+ \bar{ \widehat d}^{(i)}_R \star (\rho_Q(\widehat \Phi)^\dagger
\star \widehat Q^{(j)}_L) \Big ) \right)
\end{eqnarray}
where $W^{ij}$, $G^{ij}_u$, $G^{ij}_d$ are
Yukawa couplings and
$\bar{\Phi} = i \tau_2 \Phi^*$.

\subsection{Non-minimal versions of the NCSM}

We can use the freedom in the choice of 
traces in kinetic terms for the gauge fields
to construct non-minimal versions of the NCSM.
The general form of the gauge kinetic terms is \cite{Calmet:2001na,Aschieri:2002mc}
\beq
S_{\mathrm{gauge}} = -\frac{1}{2}  \int d^4x \,
\sum_\rho {\kappa_\rho} {\rm Tr}\Big( \rho(\widehat F_{\mu \nu})
\star \rho(\widehat F^{\mu \nu})\Big),
\eeq 
where the sum is over all unitary irreducible inequivalent representations $\rho$ of
the gauge group $G$.
The freedom in the kinetic terms 
is parametrized by real coefficients $\kappa_\rho$ that are
subject to the constraints
\beq
\frac{1}{g^2_{I}} = \sum_\rho \kappa_\rho {\rm Tr}\Big( \rho(T^a_I) \rho(T^a_I)\Big),
\eeq 
where
$g_{I}$ and $T^a_I$ are the usual ``commutative'' coupling constants and 
generators of $U(1)_Y$, $SU(2)_L$, $SU(3)_C$, respectively.
Both formulas can also be written more compactly as
\beq
S_{\mathrm{gauge}} = -\frac{1}{2}  \int d^4x \,
\mbox{\bf Tr} \frac{1}{\mbox{\bf G}^2} \widehat F_{\mu \nu} \star \widehat F^{\mu \nu},
\qquad
\frac{1}{g^2_{I}} = \mbox{\bf Tr} \frac{1}{\mbox{\bf G}^2} T^a_I T^a_I,
\eeq
where the trace \mbox{\bf Tr} is again over all representations
and \mbox{\bf G} is an operator that commutes with all
generators $T^a_I$ and encodes the coupling constants.
The possibility of new parameters in gauge theories on noncommutative
spacetime is a consequence of the fact that the gauge fields
are in general valued in the enveloping algebra of the gauge
group. 

The expansion in $\theta$ is at the same time an expansion
in the momenta. The $\theta$-expanded
action can thus be interpreted as a
low energy effective action.
In such an effective low energy description
it is natural to expect that all representations that 
appear in the commutative theory
(matter multiplets and adjoint representation)
are important. 
We should therefore consider the non-minimal version of the NCSM with
non-zero coefficients $\kappa_\rho$ at least for these
representations. The new parameters in the non-minimal NCSM can be restricted
by considering GUTs on noncommutative
spacetime \cite{Aschieri:2002mc}.

\subsection{Properties of the NCSM}

The key properties of the Noncommutative Standard Model (NCSM) are:
\begin{itemize}
\item The known elementary particles can be accomodated with their correct
charges as in the original ``commutative'' Standard Model. There is no
need to introduce new fields.
\item The noncommutative Higgs field in the minimal NCSM has distinct left and 
right hyper (and colour) charges, whose sum are the regular SM charges.
This is necesarry to obtain gauge invariant Yukawa terms.
\item In versions of the NCSM that arise from NC GUTs it is more natural
to equip the neutrino (and
other left-handed
fermion fields) with left and right charges. The neutrino
can in principle couple to photons in the
presence of spacetime noncommutativity, even though its
total charge is zero.
\item Noncommutative gauge invariance implies the existence of 
many new couplings of gauge fields: Abelian gauge bosons self-interact
via a star-commutator term that resembles the self-interaction 
of non-abelian gauge bosons and we find many new interaction terms that involve gauge fields 
as a consequence of the Seiberg-Witten maps.
\item The perturbation theory is based on the free commutative action.
Assymptotic states are the plane-wave eigenstates of the free
commutative Hamiltonian. Both ordinary interaction terms and interactions
due to noncommutative effects are treated on equal footing. This makes it
particularly simple to derive Feynman rules and compute the invariant matrix
elements of fundamental processes. While there is no need to 
reinvent perturbation theory, care has to be taken nethertheless for
a time-like $\theta$-tensor to avoid problems with unitarity.
\item Violation of spacetime symmetries 
 and in particular of angular momentum conservation and discrete
symmetries like P, CP, and possibly even CPT can be induced by spacetime noncommutativity. 
This symmetry breaking is
spontaneous in the sense that it is with respect to a fixed 
$\theta$-``vacuum''. (As long as $\theta$ is also transformed as a
tensor, everything is fully covariant.) 
\end{itemize}
The physically interpretation
of these violations of conservation laws is that angular momentum (and even
energy-momentum) can be transferred to the noncommuative 
spacetime structure in much the same way as energy can be carried away
from binary stars by gravitational waves. 

\section{Standard Model forbidden processes}

A general feature of gauge theories on noncommutative spacetime is the
appearance of many new interactions including Standard Model-forbidden
processes. The origin of these new interactions is two-fold: One source 
are the star products that let abelian gauge theory on NC spacetime resemble
Yang-Mills theory with the possibility of triple and quadruple 
gauge boson vertices. The other source are the gauge fields
in the Seiberg-Witten maps for the gauge and matter fields.
These can be pictured as a cloud of gauge bosons that dress the
original `commutative' fields and that have their origin in the interaction
between gauge fields and the NC structure of spacetime.
One of the perhaps most striking 
effects and a possible
signature of spacetime noncommutativity  
is the spontaneous breaking of continuous and discrete spacetime symmetries. 

\subsection{Triple gauge boson couplings}

New anomalous triple gauge boson interactions
that are usually forbidden by Lorentz invariance, 
angular moment conservation
and Bose statistics (Yang theorem) can
arise within the framework of the non-minimal noncommutative standard model 
\cite{Behr:2002wx,Duplancic:2003hg},
and also in the alternative approach to the NCSM given in 
\cite{Chaichian:2001py}. 

\begin{figure}
\center
\includegraphics[height=6cm]{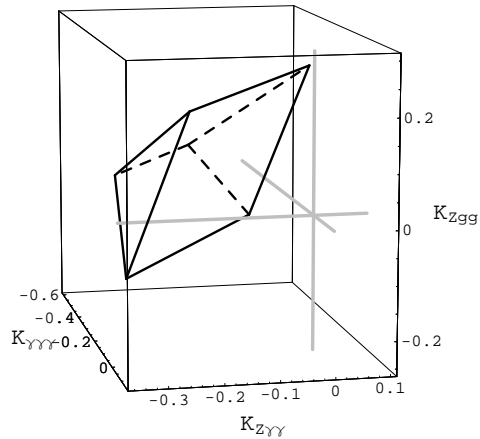}
\caption{The three-dimensional pentahedron that bounds possible values
 for the coupling constants ${\rm K}_{\gamma\gamma\gamma}$, 
 ${\rm K}_{Z\gamma\gamma}$ 
 and ${\rm K}_{Zgg}$ at the $M_Z$ scale.}
\label{simplex}       
\end{figure}

The new triple gauge boson (TGB) terms in the action have 
the following form \cite{Behr:2002wx,Duplancic:2003hg}:
\begin{eqnarray}
\lefteqn{S_{gauge}=-\frac{1}{4}\int \hspace{-1mm}d^4x\, f_{\mu \nu} f^{\mu \nu}}
 \label{action2} \\
& &\hspace{-5mm}{}
-\frac{1}{2}\int \hspace{-1mm}d^4x\, {\rm Tr}\left( F_{\mu \nu} F^{\mu \nu}\right)
-\frac{1}{2}\int\hspace{-1mm} d^4x\, {\rm Tr}\left( G_{\mu \nu} G^{\mu \nu}\right)
\nonumber \\
& &\hspace{-5mm}{}
+g_s \,\theta^{\rho\tau}\hspace{-2mm}
\int\hspace{-1mm} d^4x\, {\rm Tr}
\left(\frac{1}{4} G_{\rho \tau} G_{\mu \nu} - G_{\mu \rho} G_{\nu \tau}\right)G^{\mu \nu}\nonumber \\
& &\hspace{-5mm}{}+{g'}^3\kappa_1{\theta^{\rho\tau}}\hspace{-2mm}\int \hspace{-1mm}d^4x\,
\left(\frac{1}{4}f_{\rho\tau}f_{\mu\nu}-f_{\mu\rho}f_{\nu\tau}\right)f^{\mu\nu}
 \nonumber \\
& &\hspace{-5mm}{}+g'g^2\kappa_2 \, \theta^{\rho\tau}\hspace{-2mm}\int
\hspace{-1mm} d^4x \sum_{a=1}^{3}
\left[(\frac{1}{4}f_{\rho\tau}F^a_{\mu\nu}-
f_{\mu\rho}F^a_{\nu\tau})F^{\mu\nu,a}\!+c.p.\right]
 \nonumber \\
& &\hspace{-5mm}{}+g'g^2_s\kappa_3\, \theta^{\rho\tau}\hspace{-2mm}\int
\hspace{-1mm} d^4x \sum_{b=1}^{8}
\left[(\frac{1}{4}f_{\rho\tau}G^b_{\mu\nu}-
f_{\mu\rho}G^b_{\nu\tau})G^{\mu\nu,b}\!+c.p.\right], \nonumber 
\end{eqnarray}
where $c.p.$ means cyclic permutations.
Here $f_{\mu\nu}$, $F^a_{\mu\nu}$ and $G^b_{\mu\nu}$ are the physical field strengths corresponding 
to the groups $\rm U(1)_Y$, $\rm SU(2)_L$ and $\rm SU(3)_C$, respectively. 
The constants $\kappa_1$, $\kappa_2$ and $\kappa_3$ are functions of $1/g_i^2\; (i=1,...,6)$:
\begin{eqnarray}
\kappa_1 &=& -\frac{1}{g^2_1}-\frac{1}{4g^2_2}+\frac{8}{9g^2_3}-\frac{1}{9g^2_4}+\frac{1}{36g^2_5}
+\frac{1}{4g^2_6},
\nonumber \\
\kappa_2 &=& -\frac{1}{4g^2_2}+\frac{1}{4g^2_5}+\frac{1}{4g^2_6},
\nonumber \\
\kappa_3 &=& +\frac{1}{3g^2_3}-\frac{1}{6g^2_4}+\frac{1}{6g^2_5}.
\end{eqnarray}
The $g_i$ are the coupling constants of the non-commutative electroweak sector
up to first order in $\theta$. The appearance of new coupling constants beyond
those of the standard model reflect a freedom in the strength
of the new TGB couplings.
Matching the SM action at zeroth order in $\theta$, three consistency conditions
are imposed on (\ref{action2}):
\begin{eqnarray}
\frac{1}{{g'}^2} &=& \frac{2}{g^2_1}+\frac{1}{g^2_2}+\frac{8}{3g^2_3}+\frac{2}{3g^2_4}+\frac{1}{3g^2_5}
+\frac{1}{g^2_6},
\nonumber \\
\frac{1}{g^2}&=& \frac{1}{g^2_2}+\frac{3}{g^2_5}+\frac{1}{g^2_6},\nonumber \\
\frac{1}{g_s^2}&=& \frac{1}{g^2_3}+\frac{1}{g^2_4}+\frac{2}{g^2_5}.
\label{L2}
\end{eqnarray}
{}From the action (\ref{action2}) we extract
neutral triple-gauge boson terms which are not present in the SM Lagrangian. 
The allowed range of values for the coupling constants
\begin{eqnarray}
{\rm K}_{\gamma\gamma\gamma}&=&\frac{1}{2}\; gg'(\kappa_1 + 3 \kappa_2),
\nonumber\\
{\rm K}_{Z\gamma\gamma}&=&\frac{1}{2}\; \left[{g'}^2\kappa_1 +
\left({g'}^2-2g^2\right)\kappa_2\right],
\nonumber\\
{\rm K}_{Zgg}&=&\frac{g^2_s}{2}
\left[1+(\frac{{g'}}{g})^2\right]\kappa_3,
\nonumber
\end{eqnarray} 
compatible with conditions (\ref{L2}) and the requirement that 
$1/g_i^2 > 0$ are plotted in figure~\ref{simplex}.
The remaining three coupling constants 
${\rm K}_{Z Z \gamma}$, ${\rm K}_{Z Z Z}$ 
and ${\rm K}_{\gamma g g}$, are uniquely fixed by the equations
\begin{eqnarray}
{\rm K}_{ZZ\gamma} &=&
\frac{1}{2} \left(\frac{g}{g'}-3\frac{g'}{g}\right){\rm K}_{Z\gamma\gamma} 
-\frac{1}{2}\left(1-\frac{g^{\prime 2}}{g^2}\right){\rm K}_{\gamma\gamma\gamma},
\nonumber \\
{\rm K}_{ZZZ} &=&
\frac{3}{2} \!\left(1-\frac{g^{\prime 2}}{g^2}\right){\rm K}_{Z\gamma\gamma} 
-\frac{1}{2}\frac{g'^2}{g^2}\!\left(3-\frac{g^{\prime 2}}{g^2}\right){\rm K}_{\gamma\gamma\gamma},
\nonumber \\
{\rm K}_{\gamma gg} &=& -\frac{g}{g'}{\rm K}_{Z gg}.
\label{L7}
\end{eqnarray}
We see that any combination of two TGB coupling constants does not
vanish simultaneously due to the constraints set by the values of the SM
coupling constants at the $M_Z$ scale \cite{Duplancic:2003hg}.

We conclude that the gauge sector is a possible place for an
experimental search for noncommuative effects.
The experimental discovery of the kinematically allowed $Z\rightarrow \gamma\gamma$ decay
would indicate a violation of the Yang theorem and would be a
possible signal of spacetime non-commutativity.

\subsection{Electromagnetic properties of neutrinos}

In the presence of spacetime
noncommutativity,  neutral particles can couple to 
gauge bosons via a ${\star}$-commutator
\begin{equation}
D^{\mbox{\scriptsize NC}}_\mu \widehat \psi = \partial_\mu \widehat \psi - i e \widehat A_\mu \star \widehat \psi
+ i e \widehat\psi \star \widehat A_\mu .
\label{1}
\end{equation}
Expanding the $\star$-product in (\ref{1}) to first order in the antisymmetric
(Poisson) tensor $\theta^{\mu\nu}$, we find the following covariant derivative
on neutral spinor fields:
\begin{equation}
D^{\mbox{\scriptsize NC}}_\mu \widehat \psi = \partial_\mu \widehat \psi + 
e \theta^{\nu\rho} \, \partial_\nu\widehat A_\mu \, \partial_\rho \widehat \psi
\; .
\label{2}	
\end{equation}
We treat $\theta^{\mu\nu}$ as a constant background field of strength 
$|\theta^{\mu\nu}|=1/\Lambda^2_{\rm NC}$ that models the non-commutative structure
of spacetime in the neighborhood of the interaction region.
As $\theta$ is not invariant under Lorentz transformations, the
neutrino field can pick up angular momentum in the interaction.
The gauge-invariant action for a neutral fermion that couples to an Abelian gauge boson 
via (\ref{2}) is 
\begin{eqnarray}
S &= &\int d^4 x \,  \bar \psi 
\left[\left(\frac{}{}i\gamma^\mu \pam  - m\right) -
\frac{e}{2}F_{\mu\nu}\left(
i \theta^{\mu\nu\rho}\parh -\theta^{\mu\nu}m\right)\right]\psi\,,
\label{3} \\
{\theta}^{\mu\nu\rho} & = & {\theta}^{\mu\nu}\gamma^{\rho}+{\theta}^{\nu\rho}\gamma^{\mu}+
{\theta}^{\rho\mu}\gamma^{\nu}\,,
\nonumber
\end{eqnarray}
up to first order in $\theta$ \cite{Schupp:2002up,Minkowski:2003jg,tram}.
The noncommutative part of (\ref{3}) induces a {\it force}, proportional
to the gradient of the field strengths, which represents 
an interaction of Stern-Gerlach type \cite{sg}. 
This interaction is non-zero even for $m_\nu=0$ and in this case reduces to the
coupling between the stress--energy tensor of the neutrino $T^{\mu\nu}$ and 
the symmetric tensor composed from $\theta$ and F \cite{Schupp:2002up}.
The following is based on \cite{Minkowski:2003jg}.

\subsubsection{Neutrino dipole moments in the mass-extended Standard Model}

Following the general arguments of \cite{fy,Bernstein,shrock,mink} 
only the Dirac neutrino can have a magnetic moment. However,
the transition matrix elements relevant for $\nu_i \longrightarrow \nu_j$ may exist for both
Dirac and Majorana neutrinos.
In the neutrino-mass extended standard model \cite{mink}, the photon--neutrino effective vertex is
determined from the $\nu_i \longrightarrow \nu_j\,+\,\gamma$ transition, 
which is generated through 1-loop electroweak process that
arise from the so-called ``neutrino--penguin'' diagrams via the exchange of $\ell=e,\mu,\tau$
leptons and weak bosons, and is given by \cite{fy,tram}
\begin{eqnarray}
J_{\mu}^{\rm eff}(\gamma\nu\bar\nu)\epsilon^{\mu}(q)&=&
\left\{ F_1(q^2) {\bar \nu_j}(p')(\gamma_{\mu}q^2-q_{\mu}{\not \!q})\nu_i(p)_L 
\right. \nonumber\\
&-&\left.iF_2(q^2)\left[ m_{\nu_j}{\bar \nu_j}(p')\sigma_{\mu\nu}q^{\nu}\nu_i(p)_L 
\right.\right. \nonumber\\
&+&\left.\left. m_{\nu_i}{\bar \nu_j}(p')\sigma_{\mu\nu}q^{\nu}\nu_i(p)_R\right]\right\}
\epsilon^{\mu}(q).
\label{4}
\end{eqnarray}
The above effective interaction is invariant under the electromagnetic gauge transformation.
The first term in (\ref{4}) vanishes identically for real photon due to the electromagnetic gauge condition. 

{}From the general decomposition of the second term of the transition matrix element T (\ref{4}),
\begin{eqnarray}
{\rm T}=-i\epsilon^{\mu}(q){\bar \nu}(p')\left[A(q^2)-B(q^2)\gamma_5\right]\sigma_{\mu\nu}q^{\nu}\nu(p),
\label{5}
\end{eqnarray}
we found the following expression for the electric and magnetic dipole moments 
\begin{eqnarray}
d^{\rm el}_{ji}\equiv B(0)&=&\frac{-e}{M^{*2}}\left( m_{\nu_i}-m_{\nu_j}\right)
\sum_{\ell=e,\mu,\tau}{\rm U}^{\dagger}_{jk}{\rm U}^{}_{ki}
{\rm F}(\frac{m^2_{\ell_k}}{m^2_W}),
\label{6} \\
\mu_{ji}\equiv A(0)&=&\frac{-e}{M^{*2}}\left( m_{\nu_i}+m_{\nu_j}\right)
\sum_{\ell=e,\mu,\tau}{\rm U}^{\dagger}_{jk}{\rm U}^{}_{ki}
{\rm F}(\frac{m^2_{\ell_k}}{m^2_W}),
\label{7}
\end{eqnarray}
where $i,j,k=1,2,3$ denotes neutrino species, and  
\begin{eqnarray}
{\rm F}(\frac{m^2_{\ell_k}}{m^2_W})\simeq -\frac{3}{2}+\frac{3}{4}\frac{m^2_{\ell_k}}{m^2_W},
\;\;\;\frac{m^2_{\ell_k}}{m^2_W}\ll 1,
\label{8}
\end{eqnarray}
was obtained after the loop integration. In Eqs. (\ref{6}) and (\ref{7}) 
$M^*=4\pi\,v=3.1$ TeV, where $v=(\sqrt 2\, G_F)^{-1/2}=246$ GeV 
represents the vacuum expectation value of the scalar Higgs field \cite{DMT}.

The neutrino mixing matrix U \cite{MNS} is governing the decomposition of a coherently
produced left-handed neutrino $\widetilde{\nu}_{L,\ell}$ 
associated with charged-lepton-flavour $\ell = e, \mu, \tau$ into
the mass eigenstates $\nu_{L,i}$:
\begin{eqnarray}
|\widetilde{\nu}_{L,\ell};\,\vec p\,\rangle =
\sum_i {\rm U}_{\ell i} |\nu_{L,i};\,\vec p ,m_i\,\rangle ,
\label{9}
\end{eqnarray}
\\
For a Dirac neutrino $i=j$ \cite{Bernstein,FS}, and using 
$m_{\nu} =0.05\,{\rm eV}$ \cite{nobel}, from (\ref{7}),
in units of [$\rm e\,cm$] and Bohr magneton, we obtain 
\begin{eqnarray}
\mu_{\nu_i}&=&\frac{3e}{2M^{*2}}m_{\nu_i}\left[1-\frac{1}{2}
\sum_{\ell=e,\mu,\tau} \frac{m^2_\ell}{m^2_W}\, |{\rm U}_{\ell i}|^2\right],
\nonumber \\
{}&=&  3.0 \times 10^{-31} \,[\rm e\;cm] = 1.6\times 10^{-20} \mu_B.
\label{10}
\end{eqnarray}
{}From formula (\ref{10}) it is clear that the
chirality flip, which is necessary to induce the magnetic moment, arises only from the neutrino masses:
Dirac neutrino magnetic moment (\ref{10}) is
still much smaller than the bounds obtained from astrophysics \cite{FY,AOT}.
More detailes about Dirac neutrinos can be found in \cite{VV,VVO}.

In the case of  the off-diagonal transition moments, the first term in (\ref{8}) 
vanishes in the summation over $\ell$ due to the orthogonality condition of U (GIM cancellation)
\begin{eqnarray}
d^{\rm el}_{{\bar\nu_j}{\nu_i}}&=&\frac{3e}{2M^{*2}} 
\left( m_{\nu_i}-m_{\nu_j}\right)
\sum_{\ell=e,\mu,\tau}\frac{m^2_{\ell_k}}{m^2_W}
{\rm U}^{\dagger}_{jk}{\rm U}^{}_{ki},
\label{12}\\
{\mu}_{{\bar\nu_j}{\nu_i}}&=&\frac{3e}{2M^{*2}}
\left( m_{\nu_i}+m_{\nu_j}\right) 
\sum_{\ell=e,\mu,\tau}\frac{m^2_{\ell_k}}{m^2_W} {\rm U}^{\dagger}_{jk}{\rm U}^{}_{ki} .
\label{13}
\end{eqnarray}

In Majorana 4-component notation
the Hermitian, neutrino--flavor antisymmetric, electric and magnetic dipole operators are
\begin{eqnarray}
{D_5 \choose D}^{\mu\nu}_{ij} = 
e\,\psi^{\top}_{\rm i}\left[\rm C \;\sigma^{\mu\nu}{\gamma_5 \choose i \mbox{\bf 1}}\right]\,\psi_{\rm j}\;.
\label{14}
\end{eqnarray}
Majorana fields have the property that the particle 
is not distinguished from antiparticle. This forces us 
to use both charged lepton and antilepton propagators in the loop calculation of 
``neutrin-penguin'' diagrams. This results in a complex
antisymmetric transition matrix element T in lepton-flavour space:
\begin{eqnarray}
{\rm T_{ji}}&=&-i\epsilon^{\mu}{\bar\nu_j}
\left[(A_{ji}-A_{ij})-(B_{ji}-B_{ij})\gamma_5\right]\sigma_{\mu\nu}q^{\nu}\nu_i
\nonumber\\
&=&-i\epsilon^{\mu}{\bar\nu_j}
\left[2{\rm iIm}A_{ji}-2{\rm Re}B_{ji}\gamma_5\right]\sigma_{\mu\nu}q^{\nu}\nu_i.
\label{15}
\end{eqnarray}
{}From this equation it is explicitly clear that for $i=j$, $d^{\rm el}_{\nu_i}={\mu}_{\nu_i}=0$.
Also, considering transition moment, only one of two terms in (\ref{15}) is 
non-vanishing if the interaction respects the CP invariance, i.e. the first term vanishes if 
the relative CP of $\nu_i$ and $\nu_j$ is even, and the second term vanishes if odd \cite{shrock}.
Finally, dipole moments describing the transition from Majorana neutrino mass eigenstate-flavour 
$\nu_j$ to $\nu_k$ in the mass extended standard model reads:
\begin{eqnarray}
d^{\rm el}_{{\nu_i}{\nu_j}}&=&\frac{3e}{2M^{*2}} 
\left( m_{\nu_i}-m_{\nu_j}\right)
\sum_{\ell=e,\mu,\tau}\frac{m^2_{\ell_k}}{m^2_W}
{\rm Re}{\rm U}^{\dagger}_{jk}{\rm U}^{}_{ki},
\label{16}\\
{\mu}_{{\nu_i}{\nu_j}}&=&\frac{3e}{2M^{*2}} 
\left( m_{\nu_i}+m_{\nu_j}\right)
\sum_{\ell=e,\mu,\tau}\frac{m^2_{\ell_k}}{m^2_W} 
{\rm i\,Im}{\rm U}^{\dagger}_{jk}{\rm U}^{}_{ki} ,
\label{17}
\end{eqnarray}
For the Majorana case the neutrino-flavour mixing matrix U is approximatively unitary, i.e.  
it is necessarily of the following form \cite{DMT}
\begin{eqnarray}
\sum_{i=1}^3 {\rm U}^{\dagger}_{jk}{\rm U}^{}_{ki} = {\delta}_{ji} - \varepsilon_{ji},
\label{18}
\end{eqnarray}
where $\varepsilon$ is a hermitian nonnegative matrix (i.e. with all eigenvalues nonnegative) and 
\begin{eqnarray}
|\varepsilon|= \sqrt{{\rm Tr}\;\varepsilon^2} &=& {\cal O} \; (m_{\nu_{\rm light}}/m_{\nu_{\rm heavy}}),
\nonumber \\
&\sim& 10^{-22} \;\, {\rm to}\;\, 10^{-21}. 
\label{19}
\end{eqnarray}

For the sum and difference of neutrino masses we assume hierarchical structure 
and take $|m_3 + m_2| \simeq |m_3 - m_2| \simeq |\Delta m_{32}^2|^{1/2} = 0.05$ eV \cite{nobel}. 
For the MNS matrix elements we set $|{\rm Re}{\rm U}^{*}_{2\tau}{\rm U}^{}_{\tau 3}| 
\simeq|{\rm Im}{\rm U}^{*}_{2\tau}{\rm U}^{}_{\tau 3}| \le 0.5$.
The electric and magnetic transition dipole moments of neutrinos 
$d^{\rm el}_{\nu_2\nu_3}$ and $\mu_{\nu_2\nu_3}$ are then denoted as $\left(d^{\rm el}_{\rm mag}\right)_{23}$  
and given by
\begin{eqnarray}
\left|\left(d^{\rm el}_{\rm mag}\right)_{23}\right| &=& \;\frac{3e}{2M^{*2}}\;
\frac{m^2_{\tau}}{m^2_W}|\Delta m_{32}^2|^{1/2}
{|{\rm Re}{\rm U}^{*}_{2\tau}{\rm U}^{}_{\tau 3}|\choose |{\rm Im}{\rm U}^{*}_{2\tau}{\rm U}^{}_{\tau 3}|},
\nonumber \\
&\stackrel{<}{\sim}& 1.95 \times 10^{-30} [\rm e/eV] = 3.8 \times 10^{-35} \,[\rm e\;cm],
\nonumber \\
&=& 2.0\times 10^{-24}\, \mu_B.
\label{20}
\end{eqnarray}
The electric transition dipole moments of light neutrinos are smaller than the one of the d-quark. 
This is {\it the} order of magnitude of light neutrino transition dipole moments underlying the see--saw mechanism. 
It is by orders of magnitude smaller than in unprotected SUSY models.

\subsubsection{Limits on the noncommutativity scale}

Now we extract an upper limit on the $\star$-gradient interaction. 
The strength of the interaction (\ref{3}) becomes $|m_{\nu}\,e\,\theta \,F|$. We compare it with the dipole
interaction $|F\,\mu_{\nu_i}|$ for Dirac neutrino (\ref{10}), and with the dipole
transition interactions $|F\, d^{\rm el}_{\rm mag}|$ for Majorana case (\ref{16},\ref{17}).
Assuming that contributions from the 
neutrino-mass extended standard model are at least as large as those from noncommutativity, we
derive the following two bounds on noncommutativity arising 
from the Dirac and Majorana nature of neutrinos, respectively:
\begin{eqnarray}
\Lambda^{\rm Dirac}_{\rm NC} &\stackrel{>}{\sim}& 
\left|{\frac{e\,m_{\nu}}{\mu_{\nu}}}\right|^{1/2}\,
\simeq 1.7 \,{\rm TeV}.
\label{21}\\
\Lambda^{\rm Majorana}_{\rm NC} &\stackrel{>}{\sim}& 
\left|{\frac{e\,m_{\nu}}{\left(d^{\rm el}_{\rm mag}\right)_{23}}}\right|^{1/2}\,
\simeq 150 \,{\rm TeV}.
\label{22}
\end{eqnarray}
The fact that the neutrino mass extended standard model, as a consequence of (\ref{8}), 
produces very different
dipole moments for Dirac neutrinos (\ref{10}) and Majorana neutrinos (\ref{20}) respectively,
manifests in two different scales of noncommutativity (\ref{21}) and (\ref{22}). 
The $(m^2_{\ell}/m^2_W)$ suppression of Majorana dipole 
moments (\ref{20}) relative to the Dirac ones (\ref{10}), is the main 
source for the different scales of noncommutativity. 
The bounds on noncommutativity thus obtained fix the scale $\Lambda_{\rm NC}$
at which the expected values of the neutrino electromagnetic dipole moments 
due to noncommutativity matches the standard model contributions.

\end{document}